\documentclass[superscriptaddress,
 reprint,
 amsmath,amssymb,
 aps,color
]{revtex4-2}

\usepackage{graphicx}
\usepackage{dcolumn}
\usepackage{bm}
\usepackage{siunitx}
\usepackage[normalem]{ulem}
\usepackage[T1]{fontenc}

\begin{document}

\title{Giant critical response in a driven-dissipative quantum gas} 

\author{Ross C. Schofield}
\email{ross.schofield@imperial.ac.uk}
\author{Daniel Lim}
\affiliation{%
Blackett Laboratory, Imperial College London, Prince Consort Road, SW7 2AZ, London, United Kingdom 
}%
\author{Himadri S. Dhar}
\affiliation{Department of Physics, Indian Institute of Technology, Bombay, Powai, Mumbai 400076, India}
\author{Robert A. Nyman}
\affiliation{Mode Labs Ltd., 30 Upper High Street, Thame, Oxfordshire, England, OX9 3EZ}
\author{Akshay K. Verma}
\author{Edmund Clarke}
\author{Jon Heffernan}
\affiliation{EPSRC National Centre for III-V Technologies, University of Sheffield, S1 3JD, UK}
\author{Florian Mintert}
\author{Rupert F. Oulton}
\affiliation{%
Blackett Laboratory, Imperial College London, Prince Consort Road, SW7 2AZ, London, United Kingdom 
}%

\date{\today}

\begin{abstract}
Systems close to a phase transition turn weak perturbations into large responses. At equilibrium, this amplification is closely linked to criticality: fluctuations grow, dynamics slow, and a common soft mode controls the response. Whether this correspondence survives in driven-dissipative quantum systems, sustained by continuous pumping and loss away from thermal equilibrium, remains an open question. Here we show experimentally that it does. In a room-temperature semiconductor photon Bose-Einstein condensate, the critical slowing of spontaneous intensity fluctuations and the amplification of weak pump perturbations are measured independently. Both peak at the same condensate population, $\bar{n}_c = 1250$, where the dimensionless slowing factor and susceptibility reach the same value, $\bar{n}_c/2 = 625$. A single weakly damped collective photon-reservoir mode governs both effects. This fluctuation-response correspondence in a finite open quantum gas establishes critical susceptibility as a measurable dynamical signature of condensation, with peak gain set by system size.

\end{abstract}

\maketitle

Equilibrium phase transitions have universal features that occur independently of the system's microscopic properties:
restoring forces weaken, fluctuations grow, and weak perturbations can drive large responses~\cite{Hohenberg1977,DiCandia2023}.
This principle of inherent instability  is exemplified
by critical opalescence of mixed liquid--gas phases, where density fluctuations strongly scatter light~\cite{stanley1971introduction,hantz2021liquid}, and by single-photon detectors that exploit the superconducting phase transition~\cite{goltsman2001picosecond,hao2024compact,irwin1995application,guo2025high}. The concept also motivates quantum-critical metrology in spin systems and ultracold gases, where enhanced susceptibilities near many-body phase transitions are a resource for sensing~\cite{zanardi2006ground,garbe2020critical,yang2023quantum,difresco2024metrology}. The theoretical framework describing these phenomena, however, rests on two idealisations that are rarely satisfied in practice: thermal equilibrium and the thermodynamic limit. Real quantum systems are inherently open, continuously exchanging energy and particles with their surroundings, and are often realized at small scales where size effects are unavoidable. How genuine critical behaviour emerges in finite-sized, open, and non-equilibrium systems remains a profound and largely open question.

Driven-dissipative quantum systems bring this question into experimental reach. In contrast to systems weakly coupled to a thermal reservoir through perturbative exchange, driven-dissipative systems require continuous replenishment of particles to exist at all. They can nevertheless exhibit macroscopic order and phase-transition-like behaviour. However, the same reservoirs, gain saturation, and dissipation that sustain the steady state can also fundamentally reshape collective modes, fluctuation dynamics, and response functions~\cite{Szymanska2006,Wouters2007,Carusotto2012,Sieberer2016}. A driven-dissipative condensate therefore provides an ideal system to test whether spontaneous fluctuations and driven susceptibility remain governed by the same collective dynamics beyond equilibrium.

Condensates of light show a transition to a macroscopic quantum state of light in a driven–dissipative setting that nevertheless exhibits equilibrium-like behaviour. Although numerous systems exhibit condensation of light~\cite{Deng2002,Kasprzak2006,daskalakis2014nonlinear,Bloch2022,Klaers2010Condensation,Schofield2024,Pieczarka2024}, photon Bose-Einstein Condensates (BECs) are particularly attractive due to their demonstrated equilibrium-like characteristics at room temperature through thermalisation~\cite{Klaers2010Condensation,Schofield2024,Pieczarka2024,Kirton2013,Kirton2015}, grand canonical fluctuations~\cite{Schmitt2014,Emreztrk2023}, continuous wave operation~\cite{Schofield2024,Pieczarka2024}, and controllable system size~\cite{Walker2018,PhysRevA.98.013810}. More recently, inorganic semiconductor implementations have allowed electrically driven and continuously sustained photon condensates, with which to explore this physics in new ways~\cite{Schofield2024,Pieczarka2024,Barland2021}. 

\begin{figure*}
    \centering
    \includegraphics[width=0.8\linewidth]{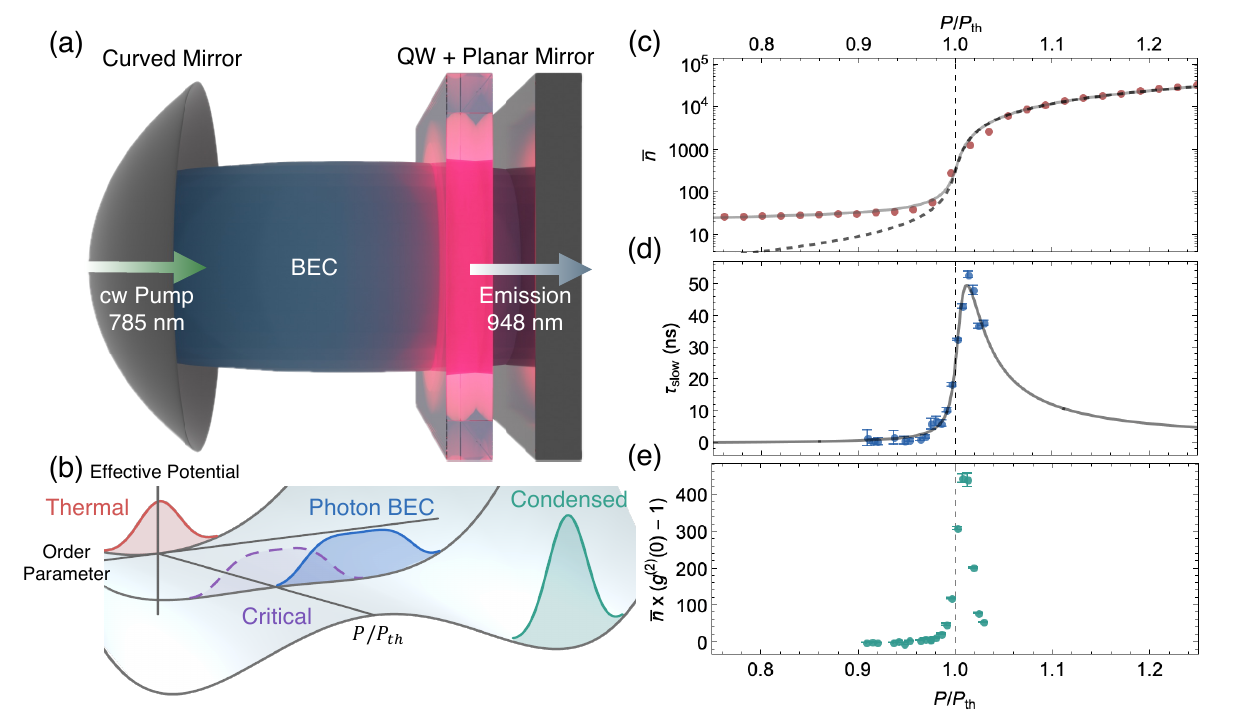}
    \caption{\textbf{Experimental platform and signatures of critical slowing near condensation.}
    (a) Schematic of the planar–spherical semiconductor microcavity, comprising a single InGaAs quantum well embedded in the planar half-cavity. The cavity is pumped continuously at 785 nm and emits at 948 nm. (b) Sketch of the free energy surface versus order parameter (condensate electric field) and normalised excitation, for a Landau laser-model. This shows how weak restoring forces emerge at the critical point leading to diverging fluctuations and their lifetime.
    (c) Measured photon number as a function of normalized excitation, $P/P_{th}$, fitted with a rate-equation model, including: total measured photons (solid line); and condensate-mode photon number, $\bar n$, (dashed line). (d) Intensity correlation time $\tau_{slow}$ extracted from fits to $g^{(2)}(\tau)$ as a function of $P/P_{th}$ showing pronounced critical slowing, fitted with Eq.~(2). (e) The Fano excess, or Mandel Q parameter, as a function of $P/P_{th}$, showing a peak in fluctuation magnitude. We note that data in (d) and (e) are limited to where $g^{(2)}(0) > 1$ is measurable. 
    }
    \label{fig:slowing}
\end{figure*}
Here we show that a finite, continuously driven semiconductor photon condensate develops a giant critical response as its photons condense. We measure critical slowing directly from the condensate's intrinsic intensity fluctuations, while independently assessing the amplification of weak excitation perturbations. Both signatures peak at the same critical condensate population, $\bar{n}_c \approx 1250$, with the same dimensionless enhancement factor, $\bar{n}_c/2$. The direct proportionality with system size stems from a dynamical model, yet captures the expected phase transition physics. This establishes a direct link between fluctuations and susceptibility at the phase boundary through a common photon-reservoir mode in a driven-dissipative quantum gas and gives a simple design rule for amplified optical transduction: the maximum gain is set by half the critical population.

We study a photon Bose-Einstein condensate in an open semiconductor microcavity comprising a planar distributed Bragg reflector (DBR) half-cavity containing a single InGaAs quantum well and a concave dielectric mirror, as shown in Fig.~\ref{fig:slowing}(a). The cavity is operated on the 15th longitudinal mode near \SI{948}{nm} and pumped continuously at \SI{785}{nm}. Pump absorption in the GaAs generates carriers that relax into the quantum well and thermalise with the photons through recombination and reabsorption, enabling room-temperature Bose-Einstein condensation of photons~\cite{Schofield2024,Figueiredo2026,Pieczarka2024,LoirettePelous2023}~(Methods). Emitted light is spectrally filtered and analysed using second-order intensity correlations $g^{(2)}(\tau)$ and pump-modulation spectroscopy (Methods), giving direct access to both spontaneous fluctuation dynamics and driven response.

We parameterise the operating point of the photon BEC using the normalised excitation $P/P_{th}$ and the condensate occupation $\bar n$. While the experimental control variable is $P/P_{th}$, $\bar n$ is the natural state variable for comparison with our dynamical model, being the square magnitude of the system's order parameter. Figures~\ref{fig:slowing}(b-e) summarise the physics of the transition as a function of the experimental control parameter $P/P_{th}$. A simple Landau laser-model of free energy as a function of order parameter and normalised excitation illustrates how large system fluctuations emerge due to reduced restoring forces at the laser phase transition (Fig~\ref{fig:slowing}(b)). The growth of $\bar n$ with excitation $P/P_{th}$ is shown in Fig.~\ref{fig:slowing}(c) and fits closely with a steady-state rate-equation model. Unlike the Landau model, $\bar n$ is non-zero at the critical point, which is a consequence of the finite system size. Meanwhile, like the Landau model, growth in $\bar n$ at the critical point is accompanied by greater intensity fluctuations that persist for longer (Figs.~\ref{fig:slowing}(d,e)).

\begin{figure}
    \centering
    \includegraphics[width=0.95\linewidth]{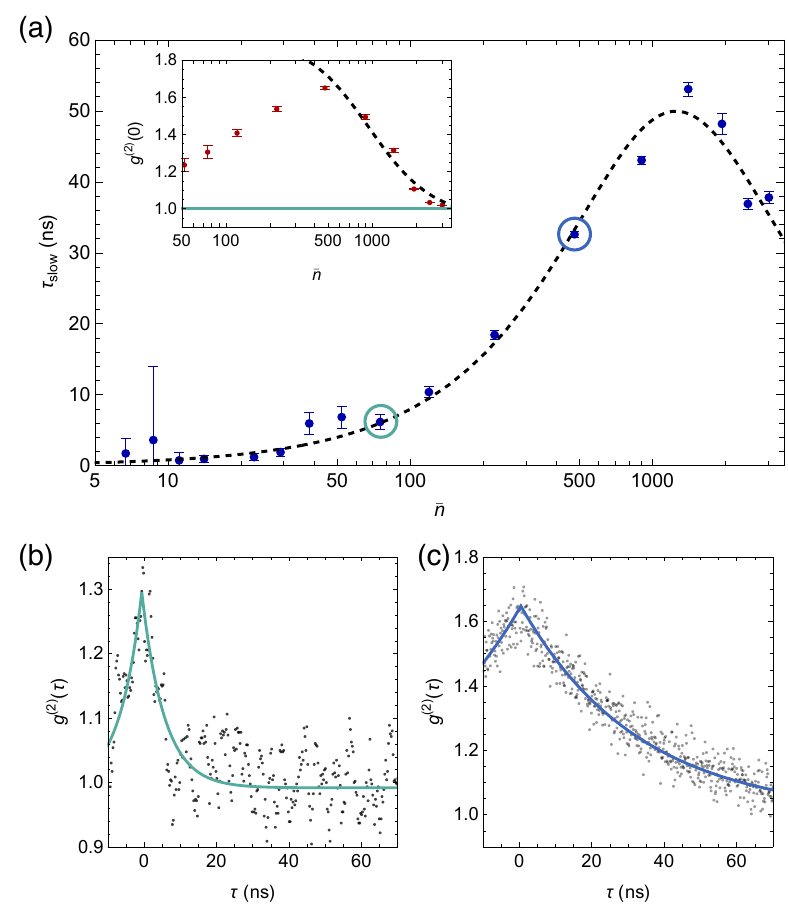}
    \caption{\textbf{Critical slowing in intensity fluctuations.} (a)~Extracted intensity-fluctuation relaxation time $\tau_{\mathrm{slow}}$ as a function of condensate occupation $\bar{n}$, showing pronounced slowing in a narrow mesoscopic window. Dashed line: Eq.~\ref{eq:tauslow_pheno}.  Inset: measured $g^{(2)}(0)$ versus $\bar{n}$ (note that uncorrected uncorrelated background/multimode contributions can render $g^{(2)}(0)$ non-monotonic; see Methods), overlaid with a phenomenological noise model. (b-c) Representative $g^{(2)}(\tau)$ traces across the transition at $\bar{n}\simeq 100$ (b) and $600$ (c) corresponding to circled measurements in (a), illustrating the evolution from thermal bunching toward Poissonian statistics and the emergence of a long correlation time. Solid line: Eq.~\ref{eq:g2fit}. }
    \label{fig:slowingg2}
\end{figure}

Figure~\ref{fig:slowingg2}(a) shows the relaxation time of the intensity fluctuations  $\tau_{\mathrm{slow}}$ as a function of $\bar n$. These values are obtained from measurements of the second-order intensity correlation function $g^{(2)}(\tau)$, representative traces of which are shown in Fig.~\ref{fig:slowingg2}(b). In the grand-canonical regime where $g^{(2)}(0)>1$ remains resolvable, we fit
\begin{equation}
g^{(2)}(\tau)\approx 1+a_{\mathrm{b}}\,e^{-|\tau-\tau_0|/\tau_{\mathrm{slow}}},
\label{eq:g2fit}
\end{equation}
where $a_{\mathrm{b}}$ quantifies the bunching amplitude, $\tau_0$ accounts for the fixed channel delay, and $\tau_{\mathrm{slow}}$ is the fluctuation relaxation time (Methods).

As in Fig.~\ref{fig:slowing}(d), pronounced slowing peaks at $\tau_{\mathrm{slow}}\approx$~\SI{50}{ns}, which is around $250$ times longer than the bare cavity lifetime ($\tau_c=\kappa^{-1}=200\pm40$~ps). It also exceeds the reservoir recombination time ($\tau_{\mathrm{rad}}=A^{-1}=0.7\pm0.1$~ns). This behaviour confirms that a collective relaxation mode of the coupled photon-reservoir becomes weakly damped near the critical point. As the emission becomes dominated by the ground state mode, $g^{(2)}(0)\mapsto 1$, as shown in the inset of Fig.~\ref{fig:slowingg2}(a), consistent with near-Poissonian statistics of the canonical regime~\cite{Schmitt2014,Emreztrk2023,Schofield2026}.

To describe the observed slowing we employ a Langevin model for the coupled evolution of photon number $\bar{n}$ and reservoir carrier population $N$. Linearising around the steady state yields a Jacobian whose eigenvalues set the relaxation rates. Using a linear-gain approximation to describe the quantum well, we find the lifetime of photon fluctuations,
\begin{equation}
\tau_{\mathrm{slow}}(\bar n)=
\frac{\kappa\tau_{\mathrm{rad}}\gamma_T+\bar n+\beta \bar n^2}
{\kappa\left(\gamma_T+\beta \bar n + \beta \bar n^2\right)},
\label{eq:tauslow_pheno}
\end{equation}
where $\tau_{\mathrm{rad}}=A^{-1}$ is the reservoir recombination time, $\beta$ is the ground-state spontaneous-emission coupling factor \cite{Nyman2018}, $\kappa$ is the cavity loss rate, and $\gamma_T=1+\alpha/\kappa$ is a dimensionless thermalisation parameter quantifying the ratio of ground-state absorption $\alpha$ to cavity loss. Equation~\eqref{eq:tauslow_pheno} captures the observed non-monotonic slowing well. $\tau_{\mathrm{slow}}$ increases on approach to the condensation critical point and decreases again as growing stimulated emission suppresses fluctuations at higher occupation, as shown by the solid curve in Fig.~\ref{fig:slowingg2}(a). Fitting Eq.~\eqref{eq:tauslow_pheno} to the measured $\tau_{\mathrm{slow}}(\bar n)$ yields $\beta=1.7\pm0.2\times10^{-6}$ and $\gamma_T=2.4\pm0.3$. 

\begin{figure}
    \centering
    \includegraphics[width=0.92\linewidth]{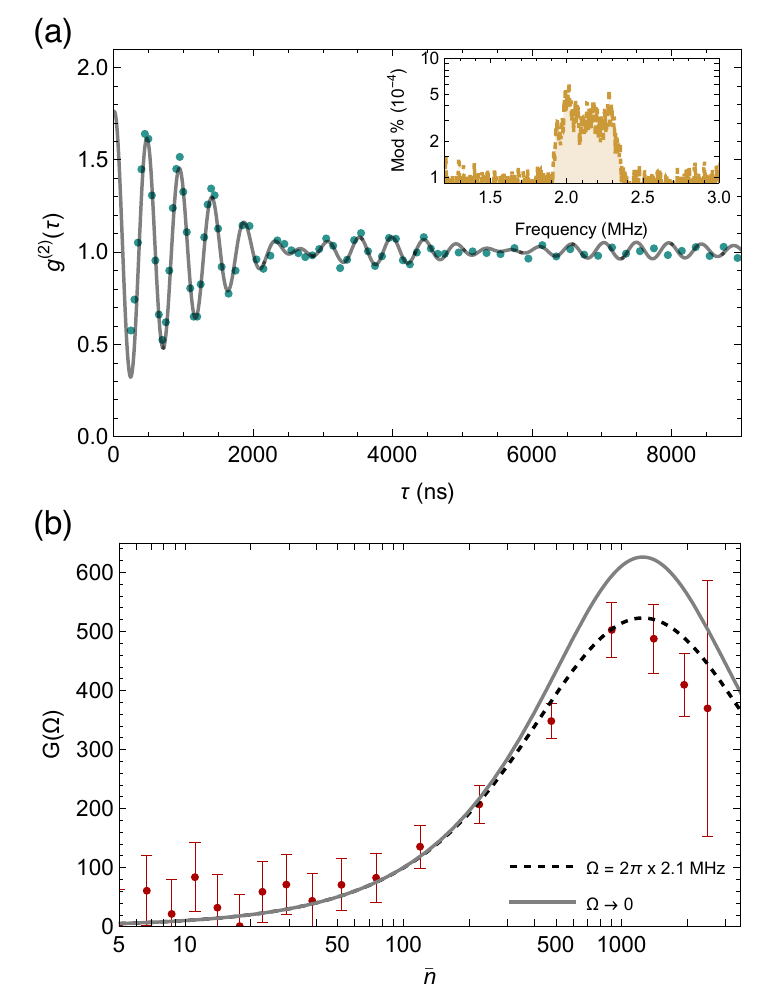}
    \caption{\textbf{Susceptibility enhancement near threshold.}
    (a) Representative long-time $g^{(2)}(\tau)$ trace measured near the critical region, showing oscillations induced by a weak narrowband intensity modulation of the pump at \SI{2.1}{MHz}. Inset: pump-noise spectrum measured with a fast photodiode, showing the corresponding spectral feature.
    (b) Measured dynamical gain $G(\Omega)$ at \SI{2.1}{MHz} versus $\bar{n}$, showing large amplification in a narrow mesoscopic window and rapid collapse beyond. Dashed line: model prediction from Eqs.~\eqref{eq:gain_rolloff} and \eqref{eq:chi_static} for $\Omega=2\pi\times\SI{2.1}{MHz}$ using parameters extracted from Fig.~\ref{fig:slowingg2}(a). Solid line: quasi-static susceptibility $\chi(\bar n)=G(\Omega\rightarrow 0)$.}
    \label{fig:suscept}
\end{figure}

We next study how the slow condensate dynamics near the critical point affects the system susceptibility to a driven excitation modulation.  Near the critical point, we observe microsecond-scale oscillations in $g^{(2)}(\tau)$, as shown in Fig.~\ref{fig:suscept}(a). These oscillations are much slower than the intrinsic critical-slowing time $\tau_{\mathrm{slow}}\lesssim$~\SI{50}{ns}, indicating that their period is set not by the condensate dynamics themselves but by an external perturbation. The oscillation frequency matches a weak intensity-noise feature of the excitation source around \SI{2.1}{MHz}, visible in the fast-photodiode spectrum shown in the inset of Fig.~\ref{fig:suscept}(a) (see Methods). The fractional modulation of the excitation is $\Delta P/P = 1.5\pm0.1\times10^{-3}$, whereas the corresponding condensate modulation can approach $\Delta\bar n/\bar n\sim 1$. Thus, weak excitation fluctuations cause a macroscopic modulation of the condensate.

To quantify this critical amplification, we define the dimensionless dynamical gain
\begin{equation}
G(\Omega)\equiv \frac{\Delta \bar{n}(\Omega)}{\bar{n}}\frac{P}{\Delta P(\Omega)},
\end{equation}
where $\Delta P(\Omega)/P$ is the fractional modulation of the pump at frequency $\Omega$, and $\Delta \bar{n}(\Omega)/\bar{n}$ is the corresponding fractional modulation of the condensate occupation inferred from the oscillatory component of $g^{(2)}(\tau)$ (Methods). Figure~\ref{fig:suscept}(b) shows $G(\Omega)$ at \SI{2.1}{MHz} as a function of $\bar n$, revealing a gain of $G\sim 500$ in a narrow excitation range and a rapid collapse either side of the critical photon number.

Within the dynamical model, a small sinusoidal modulation of the excitation results in a frequency-dependent normalised gain
\begin{equation}
G(\Omega)\approx
\frac{\chi(\bar{n})}{\sqrt{(1-\Omega^2/\Omega_{ro}^2)^2+(\Omega\tau_{\mathrm{slow}})^2}},
\label{eq:gain_rolloff}
\end{equation}
where $\chi(\bar{n})\equiv G(\Omega\!\to\!0)$ is the quasi-static susceptibility and $\Omega_{ro}=\sqrt{\beta A\kappa \bar{n} \left(1 +\gamma_T/\beta\bar{n}^2 \right)}$ is the relaxation-oscillation angular frequency. In the present regime, $\Omega\ll\Omega_{ro}$, the finite-frequency reduction is governed primarily by the factor $\Omega\tau_{\mathrm{slow}}$. Using the same linear-gain model, we obtain the closed expression
\begin{equation}
\chi(\bar n)\simeq \frac{\bar n\left(\gamma_T+\beta\bar n\right)}{\gamma_T+\beta \bar n+\beta \bar n^2}.
\label{eq:chi_static}
\end{equation}
The comparison in Fig.~\ref{fig:suscept}(b) shows that Eqs.~\eqref{eq:gain_rolloff} and \eqref{eq:chi_static}, evaluated using the fitted $\beta$ and $\gamma_T$ obtained from the critical-slowing analysis using Eq.~\ref{eq:tauslow_pheno}, reproduce the measured gain well. At the maximum slowing, $\tau_{\mathrm{slow}}\approx$~\SI{50}{ns}, and for $\Omega=2\pi\times\SI{2.1}{MHz}$ we obtain $\Omega\tau_{\mathrm{slow}}\approx0.66$, so Eq.~\eqref{eq:gain_rolloff} predicts a reduction of about $17\%$ relative to the quasi-static susceptibility, consistent with the data.

\begin{figure}
    \centering
    \includegraphics[width=0.92\linewidth]{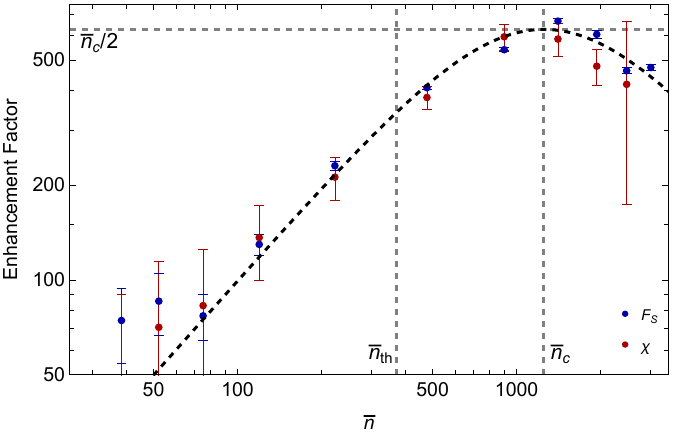}
    \caption{\textbf{Common enhancement factor from spontaneous and driven measurements.} Dimensionless enhancement factor extracted from fluctuation dynamics, $F_s=(\alpha+\kappa)\tau_{\mathrm{slow}}$ (blue), and quasi-static susceptibility, $\chi$ (red), inferred from the measured gain $G(2\pi\times\SI{2.1}{MHz})$ after correcting for the finite-frequency roll-off using Eq.~\eqref{eq:gain_rolloff}. Both quantities peak at the same critical occupation, $\bar{n}_c\approx 1.2\times 10^3$, where $\chi_c\simeq \bar{n}_c/2$. Dashed line is Eq.~\eqref{eq:chi_static}, $(\alpha + \kappa)\times$Eq.~\ref{eq:tauslow_pheno} agrees.}
    \label{fig:compare}
\end{figure}

The critical slowing and the enhanced driven response are two signatures of the same condensate dynamics near the phase boundary. Figure~\ref{fig:compare} compares the corresponding dimensionless enhancement factors and shows that they peak at the same condensate occupation. We define the total photon-loss rate of the ground state mode well below the critical point as $\alpha+\kappa = \kappa \gamma_T$, with a dimensionless slowing factor, $F_s=\kappa \gamma_T\tau_{\mathrm{slow}}$. We compare $F_s$ and the quasi-static susceptibility $\chi$ as functions of condensate occupation $\bar n$. The close agreement between $F_s$ and $\chi$ in both experiment and theory reflects a fluctuation-response relation for this driven-dissipative system. The bunching feature in $g^{(2)}(\tau)$ is the system's fast response to intrinsic noise, whilst the susceptibility is the system's response to external modulation, with both being governed by the slow condensate-reservoir mode. Their common maximum therefore provides a dynamical signature of the condensation boundary, while the finite width reflects the system's finite-size, set by $\beta$.

\begin{figure*}
    \centering
    \includegraphics[width=\linewidth]{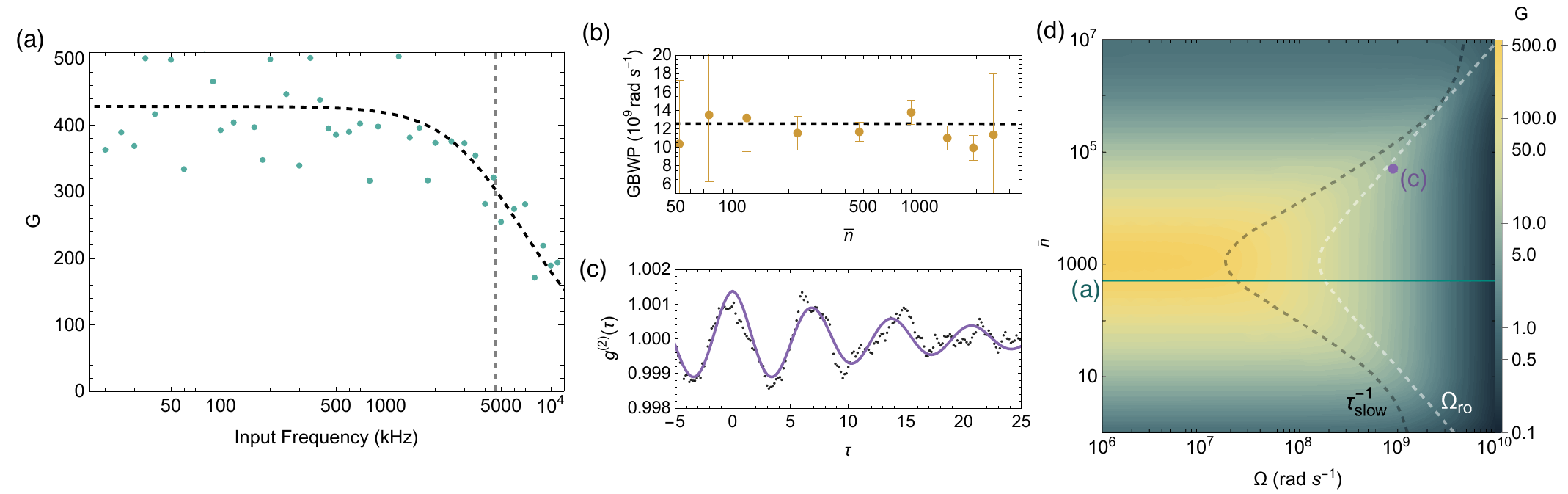}
    \caption{\textbf{Bandwidth of the critical response and relaxation oscillations.} (a) Measured dynamical gain $G(\Omega)$ as a function of modulation frequency at $\bar{n}\approx550$, showing broadband amplification with a high-frequency roll-off. Dashed line: Eq.~\eqref{eq:gain_rolloff}. Vertical line: expected corner frequency from $\tau_{\mathrm{slow}}=35\pm5$~ns. (b) Gain-bandwidth product (GBWP) versus $\bar{n}$, showing a plateau across the condensation boundary. Dashed line: theory prediction. (c) $g^{(2)}(\tau)$ measured at $\bar{n} \approx 5 \times10^4$, showing relaxation oscillations. The data are fit with a damped cosine. (d) Theory map of the gain $G$ as a function of modulation frequency $\Omega$ and condensate occupation $\bar{n}$. The horizontal teal line indicates the measurement in (a), the dashed black line the bandwidth of the critical response, and the dashed white line the relaxation-oscillation branch. The purple marker indicates the location of (c). The map shows that the strong transduction reported here arises from a slow ground-state response near the condensation boundary, whereas relaxation oscillations form a distinct underdamped branch at higher occupation.}
    \label{fig:placeholder2}
\end{figure*}

The peak enhancement occurs for a critical photon number,
\begin{equation}
    \bar{n}_c \simeq \sqrt{\frac{\gamma_T}{\beta}},
    \label{eq:nc_main}
\end{equation}
which locates the phase boundary in terms of the thermalisation strength $\gamma_T$ and the effective system size $\beta^{-1}$~\cite{Rice1994}. At the maximum enhancement point the two dominant restoring terms in the denominator of Eq.~(5) are equal, $\beta \bar n_c^2=\gamma_T$. For $\bar n_c\gg1$, the smaller term $\beta\bar n_c$ can be neglected, giving
\begin{equation}
\chi(\bar{n}_c)=\chi_c \simeq\frac{\bar n_c\gamma_T}{2\gamma_T}=\frac{\bar n_c}{2}.
\end{equation}
The factor of one-half follows from the equal contributions of thermalisation and stimulated-emission saturation to the restoring force of the intensity fluctuations at the critical point. Since $\beta$ decreases with increasing system size~\cite{vanExter1996SpontaneousBeta}, larger systems exhibit a larger critical population and a correspondingly larger peak susceptibility.

Critical amplification comes with a bandwidth cost, as the slowing of the spontaneous fluctuations also limits the driven response. To test this directly, we measure the frequency-dependent gain $G(\Omega)$ by applying controlled pump modulation from \SI{20}{kHz} to \SI{11}{MHz} at $\bar n\approx 550$. The spectrum in Fig.~\ref{fig:placeholder2}(a) exhibits a single-pole roll-off described by Eq.~\eqref{eq:gain_rolloff} in the regime $\Omega\ll\Omega_{ro}$. The fitted corner frequency corresponds to a characteristic time that agrees with $\tau_{\mathrm{slow}}=35\pm5$~ns extracted independently from the $g^{(2)}(\tau)$ bunching envelope, showing that the same weakly damped photon-reservoir mode governs both the spontaneous and driven response. 

At the condensation boundary, the increase in gain is accompanied by a decrease in bandwidth such that the product of critical gain, $\chi_c$ and bandwidth $\mathcal{B}_c$ remains approximately constant,\begin{equation}
\chi_c\,\mathcal{B}_c \simeq \kappa\gamma_T = \kappa+\alpha .
\end{equation}
Figure~\ref{fig:placeholder2}(b) shows this plateau across the boundary, in agreement with the model. The enhanced response near criticality is therefore broadband but bandwidth-limited, rather than resonantly enhanced.

This critical response is distinct from relaxation oscillations. Far above threshold, at $\bar n \approx 5\times10^4$, we observe weak oscillations in $g^{(2)}(\tau)$, shown in Fig.~\ref{fig:placeholder2}(c), with frequency consistent with $\Omega_{ro}^2 \approx \beta\kappa A\bar n$. Figure~\ref{fig:placeholder2}(d) plots $G(\Omega)$ (Eq.~\ref{eq:gain_rolloff}) and places both regimes within a single dynamical map. Near the condensation boundary at $\bar{n}_c\approx1250$, the response is dominated by the broad low-frequency ridge seen in Fig.~\ref{fig:placeholder2}(a). At higher (and lower) occupation, a separate relaxation-oscillation branch is visible. The marker extracted from Fig.~\ref{fig:placeholder2}(c) lies on this branch, showing that the same model captures both the critical regime and the far-above-threshold underdamped regime~\cite{ErglisPhotonBECRO}.

The model also provides simple design principles for controlling susceptibility and bandwidth. The peak susceptibility scales as $\chi_c\simeq \bar n_c/2$ with $\bar n_c\simeq \sqrt{\gamma_T/\beta}$, so stronger thermalisation or smaller $\beta$ shifts the peak response to larger condensate occupation and increases the susceptibility. At the same time, the approximately fixed gain-bandwidth product, $\chi_c\mathcal{B}_c \simeq \kappa\gamma_T = \kappa+\alpha$, implies that increasing bandwidth requires a larger total loss, provided thermalisation remains sufficient for condensation.

We have shown that the condensation boundary of a continuous-wave, room-temperature semiconductor photon condensate supports a strongly enhanced, but bandwidth-limited, critical dynamical response. Near the boundary, the bunching dynamics in $g^{(2)}(\tau)$ reveal relaxation times up to $\sim 50$ ns, corresponding to a dimensionless slowing factor of $\sim 625$, while weak pump modulation produces a susceptibility of the same magnitude in the same operating window. These two independent measurements show that spontaneous intensity fluctuations and driven response are governed by the same slow condensate-reservoir mode, establishing a direct fluctuation-response correspondence and distinguishing the observed enhancement from conventional relaxation oscillations.

A minimal model captures the steady-state operating point, the critical slowing, the susceptibility peak, and the finite-frequency roll-off within a common framework. This correspondence emerges at a critical population of only $\bar n_c\simeq1250$, where finite-size rounding set by the effective system size parameter $\beta^{-1}$ is directly observable. This demonstrates that critical physics survives well outside the thermodynamic limit. These results establish critical susceptibility as a measurable dynamical consequence of photon condensation and identify semiconductor photon condensates as an experimentally accessible platform for susceptibility-enhanced optical transduction in driven-dissipative many-body systems.

\bibliographystyle{apsrev4-2}
\bibliography{export.bib}

@article{Szymanska2006,
   author = {M. H. Szymanska and J. Keeling and P. B. Littlewood},
   doi = {10.1103/PhysRevLett.96.230602},
   issue = {23},
   journal = {Physical Review Letters},
   month = {3},
   title = {Non-equilibrium quantum condensation in an incoherently pumped dissipative system},
   volume = {96},
   url = {http://arxiv.org/abs/cond-mat/0603447 http://dx.doi.org/10.1103/PhysRevLett.96.230602},
   year = {2006}
}

@article{Wouters2007,
   author = {Michiel Wouters and Iacopo Carusotto},
   doi = {10.1103/PhysRevLett.99.140402},
   issn = {00319007},
   issue = {14},
   journal = {Physical Review Letters},
   month = {10},
   pages = {140402},
   publisher = {American Physical Society},
   title = {Excitations in a Nonequilibrium Bose-Einstein Condensate of Exciton Polaritons},
   volume = {99},
   url = {https://journals.aps.org/prl/abstract/10.1103/PhysRevLett.99.140402},
   year = {2007}
}

@article{Sieberer2016,
   author = {L. M. Sieberer and M. Buchhold and S. Diehl},
   doi = {10.1088/0034-4885/79/9/096001},
   issue = {9},
   journal = {Reports on Progress in Physics},
   month = {8},
   publisher = {Institute of Physics Publishing},
   title = {Keldysh Field Theory for Driven Open Quantum Systems},
   volume = {79},
   url = {http://arxiv.org/abs/1512.00637 http://dx.doi.org/10.1088/0034-4885/79/9/096001},
   year = {2016}
}

@article{Carusotto2012,
   author = {Iacopo Carusotto and Cristiano Ciuti},
   doi = {10.1103/RevModPhys.85.299},
   issue = {1},
   journal = {Reviews of Modern Physics},
   month = {10},
   pages = {299-366},
   publisher = {American Physical Society},
   title = {Quantum fluids of light},
   volume = {85},
   url = {http://arxiv.org/abs/1205.6500 http://dx.doi.org/10.1103/RevModPhys.85.299},
   year = {2012}
}

@book{stanley1971introduction,
  title={Introduction to Phase Transitions and Critical Phenomena},
  author={Stanley, H. Eugene},
  year={1971},
  publisher={Oxford University Press},
  address={New York and Oxford},
  series={International Series of Monographs on Physics}
}

@article{goltsman2001picosecond,
  title={Picosecond superconducting single-photon optical detector},
  author={Gol'tsman, G. N. and Okunev, O. and Chulkova, G. and Lipatov, A. and Semenov, A. and Smirnov, K. and Voronov, B. and Dzardanov, A. and Williams, C. and Sobolewski, Roman},
  journal={Applied Physics Letters},
  volume={79},
  number={6},
  pages={705--707},
  year={2001},
  publisher={American Institute of Physics},
  doi={10.1063/1.1388868}
}

@article{irwin1995application,
  title={An application of electrothermal feedback for high resolution cryogenic particle detection},
  author={Irwin, K. D.},
  journal={Applied Physics Letters},
  volume={66},
  number={15},
  pages={1998--2000},
  year={1995},
  publisher={American Institute of Physics},
  doi={10.1063/1.113674}
}

@article{zanardi2006ground,
  title={Ground state overlap and quantum phase transitions},
  author={Zanardi, Paolo and Paunkovi{\'c}, Nikola},
  journal={Physical Review E},
  volume={74},
  number={3},
  pages={031123},
  year={2006},
  publisher={American Physical Society},
  doi={10.1103/PhysRevE.74.031123}
}

@article{garbe2020critical,
  title={Critical Quantum Metrology with a Finite-Component Quantum Phase Transition},
  author={Garbe, Louis and Abah, Obinna and Felicetti, Simone and Puebla, Ricardo},
  journal={Physical Review Letters},
  volume={124},
  number={12},
  pages={120504},
  year={2020},
  publisher={American Physical Society},
  doi={10.1103/PhysRevLett.124.120504}
}

@article{hantz2021liquid,
  title={Liquid--Liquid Demonstrations: Critical Opalescence},
  author={Hantz, Kathleen L. and Ganske, Jane A.},
  journal={Journal of Chemical Education},
  volume={98},
  number={7},
  pages={2364--2369},
  year={2021},
  publisher={American Chemical Society},
  doi={10.1021/acs.jchemed.0c01518}
}

@article{difresco2024metrology,
  title={Metrology and multipartite entanglement in measurement-induced phase transition},
  author={Di Fresco, Giovanni and Spagnolo, Bernardo and Valenti, Davide and Carollo, Angelo},
  journal={Quantum},
  volume={8},
  pages={1326},
  year={2024},
  publisher={Verein zur F{\"o}rderung des Open Access Publizierens in den Quantenwissenschaften},
  doi={10.22331/q-2024-04-30-1326}
}

@article{hao2024compact,
  title={A compact multi-pixel superconducting nanowire single-photon detector array supporting gigabit space-to-ground communications},
  author={Hao, Hao and Zhao, Qing-Yuan and Huang, Yang-Hui and Deng, Jie and Yang, Fan and Ru, Sai-Ying and Liu, Zhen and Wan, Chao and Liu, Hao and Li, Zhi-Jian and others},
  journal={Light: Science \& Applications},
  volume={13},
  number={1},
  pages={25},
  year={2024},
  publisher={Nature Publishing Group},
  doi={10.1038/s41377-023-01374-1}
}

@article{guo2025high,
  title={A High-Speed, High-Resolution Transition Edge Sensor Spectrometer for Soft X-Rays at the Advanced Photon Source},
  author={Guo, S. and Quaranta, O. and Weber, J. C. and McChesney, J. L. and others},
  journal={IEEE Transactions on Applied Superconductivity},
  volume={35},
  number={3},
  pages={1--5},
  year={2025},
  doi={10.1109/TASC.2025.3526110}
}

@article{yang2023quantum,
  title={Quantum-enhanced sensing of a magnetic field gradient with a quantum phase transition},
  author={Yang, L-P. and others},
  journal={Nature Communications},
  volume={14},
  pages={5214},
  year={2023},
  publisher={Nature Publishing Group},
  doi={10.1038/s41467-023-40915-x}
}

@article{Schofield2026,
   author = {Ross C. Schofield and Daniel Lim and Nathan R. Gemmell and Edmund Clarke and Ian Farrer and Aristotelis Trapalis and Jon Heffernan and Rupert F. Oulton},
   doi = {10.1103/1zxb-jyrf},
   issn = {2331-7019},
   issue = {5},
   journal = {Physical Review Applied},
   month = {5},
   pages = {L051002},
   title = {Semiconductor photon Bose-Einstein condensate as a practical light source for range finding},
   volume = {25},
   url = {https://link.aps.org/doi/10.1103/1zxb-jyrf},
   year = {2026}
}

@article{DiCandia2023,
   author = {R. Di Candia and F. Minganti and K. V. Petrovnin and G. S. Paraoanu and S. Felicetti},
   doi = {10.1038/s41534-023-00690-z},
   issn = {20566387},
   issue = {1},
   journal = {npj Quantum Information 2023 9:1},
   month = {3},
   pages = {23-},
   publisher = {Nature Publishing Group},
   title = {Critical parametric quantum sensing},
   volume = {9},
   url = {https://www.nature.com/articles/s41534-023-00690-z},
   year = {2023}
}

@article{Hohenberg1977,
   author = {P. C. Hohenberg and B. I. Halperin},
   doi = {10.1103/RevModPhys.49.435},
   issn = {00346861},
   issue = {3},
   journal = {Reviews of Modern Physics},
   month = {7},
   pages = {435},
   publisher = {American Physical Society},
   title = {Theory of dynamic critical phenomena},
   volume = {49},
   url = {https://journals.aps.org/rmp/abstract/10.1103/RevModPhys.49.435},
   year = {1977}
}

@article{Schmitt2014,
   author = {Julian Schmitt and Tobias Damm and David Dung and Frank Vewinger and Jan Klaers and Martin Weitz},
   doi = {10.1103/PHYSREVLETT.112.030401/SUPPLEMENTAL.TEX},
   issn = {00319007},
   issue = {3},
   journal = {Physical Review Letters},
   month = {1},
   pages = {030401},
   publisher = {American Physical Society},
   title = {Observation of grand-canonical number statistics in a photon bose-einstein condensate},
   volume = {112},
   url = {https://journals.aps.org/prl/abstract/10.1103/PhysRevLett.112.030401},
   year = {2014}
}

@article{Pieczarka2024,
   author = {Maciej Pieczarka and Marcin Gębski and Aleksandra N. Piasecka and James A. Lott and Axel Pelster and Michał Wasiak and Tomasz Czyszanowski},
   doi = {10.1038/S41566-024-01478-Z;SUBJMETA},
   issn = {17494893},
   issue = {10},
   journal = {Nature Photonics},
   month = {10},
   pages = {1090-1096},
   publisher = {Nature Research},
   title = {Bose–Einstein condensation of photons in a vertical-cavity surface-emitting laser},
   volume = {18},
   url = {https://www.nature.com/articles/s41566-024-01478-z},
   year = {2024}
}

@article{Emreztrk2023,
   author = {Fahri Emre Öztürk and Frank Vewinger and Martin Weitz and Julian Schmitt},
   doi = {10.1103/PhysRevLett.130.033602},
   journal = {Physical Review Letters},
   pages = {33602},
   title = {Fluctuation-Dissipation Relation for a Bose-Einstein Condensate of Photons},
   volume = {130},
   year = {2023}
}

@article{Schofield2024,
   author = {Ross C. Schofield and Ming Fu and Edmund Clarke and Ian Farrer and Aristotelis Trapalis and Himadri S. Dhar and Rick Mukherjee and Toby Severs Millard and Jon Heffernan and Florian Mintert and Robert A. Nyman and Rupert F. Oulton},
   doi = {10.1038/s41566-024-01491-2},
   issn = {1749-4885},
   issue = {10},
   journal = {Nature Photonics},
   month = {10},
   pages = {1083-1089},
   publisher = {Nature Research},
   title = {Bose–Einstein condensation of light in a semiconductor quantum well microcavity},
   volume = {18},
   url = {https://www.nature.com/articles/s41566-024-01491-2},
   year = {2024}
}

@article{Figueiredo2026,
   author = {José L Figueiredo and Ross Schofield and Ming Fu and Robert Nyman and Rupert F Oulton and Hugo Terças and Florian Mintert},
   doi = {10.1088/1361-6633/ae65da},
   isbn = {2509.05062v1},
   issn = {0034-4885},
   journal = {Reports on Progress in Physics},
   month = {4},
   title = {A Quantum Kinetic Theory of Photon Bose-Einstein Condensation in Semiconductors},
   url = {https://iopscience.iop.org/article/10.1088/1361-6633/ae65da},
   year = {2026}
}

@article{LoirettePelous2023,
   author = {Aurelian Loirette‐Pelous and Jean‐Jacques Greffet},
   doi = {10.1002/lpor.202300366},
   issn = {1863-8880},
   issue = {11},
   journal = {Laser \& Photonics Reviews},
   month = {11},
   title = {Photon Bose–Einstein Condensation and Lasing in Semiconductor Cavities},
   volume = {17},
   url = {https://onlinelibrary.wiley.com/doi/10.1002/lpor.202300366},
   year = {2023}
}

@article{Kirton2015,
   author = {Peter Kirton and Jonathan Keeling},
   doi = {10.1103/PhysRevA.91.033826},
   issn = {1050-2947},
   issue = {3},
   journal = {Physical Review A},
   month = {3},
   pages = {033826},
   publisher = {American Physical Society},
   title = {Thermalization and breakdown of thermalization in photon condensates},
   volume = {91},
   url = {https://link.aps.org/doi/10.1103/PhysRevA.91.033826},
   year = {2015}
}

@article{Kasprzak2006,
   author = {J. Kasprzak and M. Richard and S. Kundermann and A. Baas and P. Jeambrun and J. M. J. Keeling and F. M. Marchetti and M. H. Szymańska and R. André and J. L. Staehli and V. Savona and P. B. Littlewood and B. Deveaud and Le Si Dang},
   doi = {10.1038/nature05131},
   issn = {0028-0836},
   issue = {7110},
   journal = {Nature},
   month = {9},
   pages = {409-414},
   publisher = {Nature Publishing Group},
   title = {Bose–Einstein condensation of exciton polaritons},
   volume = {443},
   url = {https://www.nature.com/articles/nature05131},
   year = {2006}
}

@article{Deng2002,
   author = {Hui Deng and Gregor Weihs and Charles Santori and Jacqueline Bloch and Yoshihisa Yamamoto},
   doi = {10.1126/science.1074464},
   issn = {0036-8075},
   issue = {5591},
   journal = {Science},
   month = {10},
   pages = {199-202},
   title = {Condensation of Semiconductor Microcavity Exciton Polaritons},
   volume = {298},
   url = {https://www.science.org/doi/10.1126/science.1074464},
   year = {2002}
}

@article{Barland2021,
   author = {S. Barland and P. Azam and G. L. Lippi and R. A. Nyman and R. Kaiser},
   doi = {10.1364/oe.409344},
   issn = {10944087},
   issue = {6},
   journal = {Optics Express},
   month = {3},
   pages = {8368},
   pmid = {33820285},
   publisher = {Optica Publishing Group},
   title = {Photon thermalization and a condensation phase transition in an electrically pumped semiconductor microresonator},
   volume = {29},
   year = {2021}
}

@article{Bloch2022,
   author = {Jacqueline Bloch and Iacopo Carusotto and Michiel Wouters},
   doi = {10.1038/s42254-022-00464-0},
   issn = {2522-5820},
   issue = {7},
   journal = {Nature Reviews Physics},
   month = {6},
   pages = {470-488},
   publisher = {Nature Publishing Group},
   title = {Non-equilibrium Bose–Einstein condensation in photonic systems},
   volume = {4},
   url = {https://www.nature.com/articles/s42254-022-00464-0},
   year = {2022}
}

@article{Walker2018,
   author = {Benjamin T. Walker and Lucas C. Flatten and Henry J. Hesten and Florian Mintert and David Hunger and Aurélien A. P. Trichet and Jason M. Smith and Robert A. Nyman},
   doi = {10.1038/s41567-018-0270-1},
   issn = {1745-2473},
   issue = {12},
   journal = {Nature Physics},
   month = {12},
   pages = {1173-1177},
   publisher = {Nature Publishing Group},
   title = {Driven-dissipative non-equilibrium Bose–Einstein condensation of less than ten photons},
   volume = {14},
   url = {https://www.nature.com/articles/s41567-018-0270-1},
   year = {2018}
}

@article{Klaers2010Condensation,
   author = {Jan Klaers and Julian Schmitt and Frank Vewinger and Martin Weitz},
   doi = {10.1038/nature09567},
   issn = {1476-4687},
   issue = {7323},
   journal = {Nature 2010 468:7323},
   month = {11},
   pages = {545-548},
   publisher = {Nature Publishing Group},
   title = {Bose–Einstein condensation of photons in an optical microcavity},
   volume = {468},
   url = {https://www.nature.com/articles/nature09567},
   year = {2010}
}

@article{Nyman2018,
   author = {Robert A. Nyman and Benjamin T. Walker},
   doi = {10.1080/09500340.2017.1404655},
   issn = {0950-0340},
   issue = {5-6},
   journal = {Journal of Modern Optics},
   month = {3},
   pages = {754-766},
   publisher = {Taylor \& Francis},
   title = {Bose-Einstein condensation of photons from the thermodynamic limit to small photon numbers},
   volume = {65},
   url = {https://www.tandfonline.com/doi/full/10.1080/09500340.2017.1404655},
   year = {2018}
}

@article{Kirton2013,
   author = {Peter Kirton and Jonathan Keeling},
   doi = {10.1103/PhysRevLett.111.100404},
   issn = {0031-9007},
   issue = {10},
   journal = {Physical Review Letters},
   month = {9},
   pages = {100404},
   title = {Nonequilibrium Model of Photon Condensation},
   volume = {111},
   url = {https://link.aps.org/doi/10.1103/PhysRevLett.111.100404},
   year = {2013}
}

@article{PhysRevA.98.013810,
  title = {Density distribution of a Bose-Einstein condensate of photons in a dye-filled microcavity},
  author = {Greveling, S. and Perrier, K. L. and van Oosten, D.},
  journal = {Phys. Rev. A},
  volume = {98},
  issue = {1},
  pages = {013810},
  numpages = {5},
  year = {2018},
  month = {Jul},
  publisher = {American Physical Society},
  doi = {10.1103/PhysRevA.98.013810},
  url = {https://link.aps.org/doi/10.1103/PhysRevA.98.013810}
}

@article{daskalakis2014nonlinear,
  title={Nonlinear interactions in an organic polariton condensate},
  author={Daskalakis, K. S. and Maier, S. A. and Murray, R. and K{\'e}na-Cohen, S.},
  journal={Nature Materials},
  volume={13},
  number={3},
  pages={271--278},
  year={2014},
  publisher={Nature Publishing Group},
  doi={10.1038/nmat3874}
}

@article{Rice1994,
  title = {Photon statistics of a cavity-QED laser: A comment on the laser--phase-transition analogy},
  author = {Rice, Perry R. and Carmichael, H. J.},
  journal = {Physical Review A},
  volume = {50},
  issue = {5},
  pages = {4318--4329},
  numpages = {0},
  year = {1994},
  month = {Nov},
  publisher = {American Physical Society},
  doi = {10.1103/PhysRevA.50.4318},
  url = {https://link.aps.org/doi/10.1103/PhysRevA.50.4318}
}

@article{vanExter1996SpontaneousBeta,
  author  = {van Exter, M. P. and Nienhuis, G. and Woerdman, J. P.},
  title   = {Two Simple Expressions for the Spontaneous Emission Factor $\beta$},
  journal = {Physical Review A},
  volume  = {54},
  number  = {4},
  pages   = {3553--3558},
  year    = {1996},
  doi     = {10.1103/PhysRevA.54.3553}
}

@article{ErglisPhotonBECRO,
  title = {Time-Periodic Driving of a Bath-Coupled Open Quantum Gas of Light},
  author = {Erglis, Andris and Sazhin, Alexander and Vewinger, Frank and Weitz, Martin and Buhmann, Stefan Yoshi and Schmitt, Julian},
  journal = {Phys. Rev. Lett.},
  volume = {135},
  issue = {3},
  pages = {033603},
  numpages = {7},
  year = {2025},
  month = {Jul},
  publisher = {American Physical Society},
  doi = {10.1103/922g-1fng},
  url = {https://link.aps.org/doi/10.1103/922g-1fng}
}

\end{document}